\documentstyle[12pt,epsf]{article}
\font\elevenbf=cmbx10 scaled\magstep 1
\font\elevenrm=cmr10 scaled\magstep 1
 1

\renewenvironment{thebibliography}[1]
 { \elevenrm
   \begin{list}{\arabic{enumi}.}
    {\usecounter{enumi} \setlength{\parsep}{0pt}
     \setlength{\itemsep}{3pt} \settowidth{\labelwidth}{#1.}
     \sloppy
    }}{\end{list}}

\begin{document}
\noindent
\thispagestyle{empty}
\begin{flushright}
\vbox{
{\bf TTP93-34}\\
{\bf December 1993}\\
 }
\end{flushright}
\begin{center}
  \begin{large}
HADRON RADIATION IN LEPTONIC $Z$ DECAYS\\
  \end{large}
  \vspace{0.8cm}
  \begin{large}
   A.H. Hoang${}^{a}$, 
   M. Je\.zabek${}^{a,}\,{}^b$, J.H. K\"uhn${}^a$ and T. Teubner${}
^{a,}$\footnote{e--mail: tt@ttpux2.physik.uni-karlsruhe.de}
 \\
  \end{large}
  \vspace{0.3cm}
${}^a$ Institut f\"ur Theoretische Teilchenphysik,Universit\"at Karlsruhe,
    76128 Karlsruhe \\
${}^b$ Institute of Nuclear Physics, Kawiory 26a, PL-30055 Cracow \\
  \vspace{1.0cm}
  {\bf Abstract}\\
\vspace{0.5cm}
\noindent
\begin{minipage}{15.0cm}
\begin{small}
The rate for the final state radiation of hadrons in leptonic
$Z$ decays is evaluated, using as input experimental data for
$\sigma (e^+e^-\to hadrons)$ in the low energy region. Configurations
with a lepton pair of large and a hadronic state of low invariant
mass are dominant. A relative rate
$\Gamma_{l\bar l had}/\Gamma_{l\bar l}=6.3\times 10^{-4}$ is calculated.
This result is about twice the prediction based on a parton model
calculation with a quark mass of $300$ MeV.
The rate for secondary production of heavy quarks is calculated in the
same formalism.
\end{small}
\end{minipage}
\end{center}
\vspace{1.5cm}
\noindent 
{\em Introduction.}
Rare decay modes of the $Z$ boson into final states with two
fermion--antifermion pairs could provide the first signal for
new physics, in particular for the production of a new scalar boson.
They are, however, also predicted in the context of the Standard Model.
A thorough understanding of production rates and distributions
is, therefore, mandatory. The final states can be sorted into three
classes: Into a class (i) with quarks only, another class (ii) which
involves a quark plus a lepton pair and finally, a class (iii) with
leptons only. Class (i) contributes in order $\alpha^2$ to the purely
hadronic final state and is part of the hadronic decay rate calculated
in \cite{Chet, Kniehl, vdB}. Class (iii) can be treated by standard methods
of perturbation theory. These formulae have also been applied to \cite{vdB}
class (ii) final states, assigning a mass of 300 MeV to the light quarks.
However, in practice the dominant contributions consist of a few pions
with low invariant masses. A treatment, based on the experimental
knowledge of $R_{had}$, the rate for hadron production through the
virtual photon, should complement and in fact superseed the purely
perturbative treatment.\\
The techniques employed in this paper have been developed \cite{KKKS}
in the context of initial state radiation. They are based on the
observation that the information relevant for the production of
low mass hadronic states through the virtual photon can be encoded
in a few moments of $R_{had}$.
For hadrons these have to be calculated numerically. For leptons, however,
they can be calculated analytically, providing a convenient test of results
that can be found in the literature.\\
The subsequent treatment will be limited to contributions induced by virtual
photons only. Contributions to the rate where the secondary fermions
are produced through neutral current interactions can safely be neglected.\\
\begin{figure}
\label{fig1}
\begin{center}
\leavevmode
\epsffile[0 70 397 213]{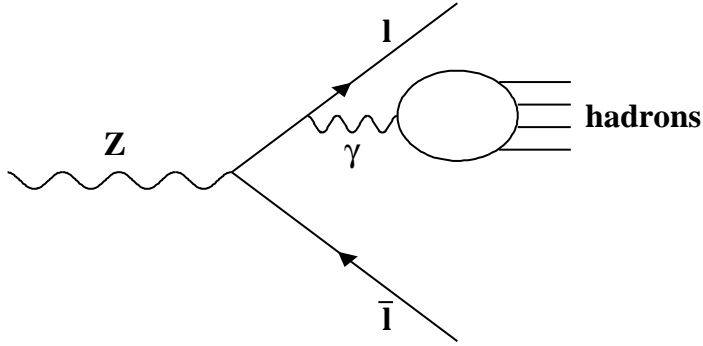}
\caption{{\em Typical diagram, describing the radiation of hadronic
states from the $l \bar l$ final state.}}
\end{center}
\end{figure}

\noindent
{\em Hadronic final state radiation.}
Neglecting the mass of leptons, the rate for $Z \to l \bar l + hadrons$ which
is induced through the amplitude depicted in Fig.1 can be cast
into the following form:
\begin{equation}\
\frac{\Gamma_{l \bar l had}}{\Gamma_{l \bar l}} = 
\frac{1}{6} \left(\frac{\alpha}{\pi}\right)^2 
\int_0^s \frac{{\rm d}s'}{s'} R(s') F(s'/s)\,,
\label{equ.1}
\end{equation}
with
\begin{eqnarray}
F(\lambda) &=& (\lambda+1)^2 \left[ -2\ln^2(\lambda+1)-4 {\rm Li}_2\left(
\frac{1}{1+\lambda}\right)+\ln^2\lambda \right] + \nonumber\\
& &(3\lambda^2+4\lambda+3) \ln\lambda + \frac{1}{3}(1+\lambda)^2\pi^2
-5(\lambda^2-1)\,,
\label{equ.2}
\end{eqnarray}
where ${\rm Li}_2$ denotes the dilogarithm
${\rm Li}_2(y)=-\int_0^y\frac{\ln(1-x)}{x}{\rm d}x$.
Neglecting trivial factors, $F(s'/s)$ gives the rate for the decay of a vector
boson of mass $s$ into a vector boson of mass $s'$ plus a pair of massless
fermions $(l \bar l\,)$. The formula is valid, as
long as the fermion mass $m_l$
is far smaller than $\sqrt{s'}$. Eq.\ref{equ.1} can be viewed as a
representation of the rate for $l \bar l + hadrons$ through a superposition
of a series of resonances with masses $s'$, distributed with weight $R(s')$.\\
Since the threshold for pion production is located at $2m_{\pi}$ and the
minimal $s'$ is far larger than $4 m_l^2$ this approach is well justified
if $l$ is an electron. The approximation can also be applied for muons, since
$R(s')$ starts to contribute significantly
for $\sqrt{s'}\stackrel{>}{{\scriptstyle\sim}}m_{\rho}$, which is significantly
larger than the muon mass. However, it is not applicable
for $\tau^+\tau^-+hadrons$. It is valid for the contribution to the rate
for $l_1 \bar l_1 l_2 \bar l_2$ where the heavier lepton is radiated off
the light leptons, and more generally, if the invariant mass of
the "secondary" lepton pair is constrained through appropriate cuts.\\
The final state under consideration consists typically of a lepton
pair with large
and a hadronic state with low invariant mass which is emitted collinear
with the lepton momentum. [The converse possibility,
where the virtual photon is radiated off the quark can be formally
treated with a similar technique. However, the final state consists
mostly of relatively soft and collinear photons of low invariant mass
and the corresponding hadronic amplitudes are difficult to tackle. Since
this configuration leads, typically, to a lepton pair of low invariant
mass, it can be easily distinguished from
\pagebreak[4]
the former configuration.]\\
The evaluation of (\ref{equ.1}) proceeds as follows. The integral can be
split into a part which is governed by the large $s'$ behaviour of
$R$ and a remainder which is sensitive to the details of the threshold
behaviour:
\begin{equation}
\int_0^s \frac{{\rm d}s'}{s'} R(s') F\left(s'/s\right) = 
R(\infty) \int_{(2\mu)^2/s}^1 \frac{{\rm d}\lambda}{\lambda} F(\lambda) +
\int_{(2\mu)^2}^s \frac{{\rm d}s'}{s'} \left[R(s') - R(\infty)\right]
F\left(s'/s\right)\,,
\label{equ.3}
\end{equation}
with
\begin{equation}
R(s) \stackrel{s \to \infty}{\longrightarrow} R(\infty) +
{\cal O}(\ln^{-3}s)\,.
\label{equ.4}
\end{equation}
At this point it is useful to observe that for $\mu$ one may take any positive
value, as long as the function $R(s)$ vanishes by definition
below $2 m_{\pi}$\,. However, for the following discussion $\mu$ will
be simply identified with $m_{\pi}$\,. The first integral can be
evaluated in a straightforward way.
\begin{eqnarray}
\int_{1/y}^1 \frac{{\rm d}\lambda}{\lambda} F(\lambda) &=&
-\frac{19}{2}+6\zeta_3+8 {\rm Li}_3(-1/y)-\frac{1}{3}\ln^3(1/y)
+\frac{6}{y} +\frac{7}{2y^2}-\nonumber\\
& &\left(5-2\zeta_2+4 {\rm Li}_2(-1/y)+\frac{6}{y}
+\frac{2}{y^2}\right) \ln(1/y)-\nonumber\\
& &\left(3+\frac{4}{y}+\frac{1}{y^2}\right)\left(-\zeta_2-2 {\rm Li}_2(-1/y)+
\frac{1}{2}\ln^2(1/y)-2\ln(1/y)\ln(1+1/y)\right)
\label{equ.9a}
\end{eqnarray}
where the trilogarithm ${\rm Li}_3$ is defined by ${\rm Li}_3(y)=\int_0^y
\frac{{\rm Li}_2(x)}{x}{\rm d}x$, $\zeta_2={\rm Li}_2(1)=\frac{\pi^2}{6}$,
$\zeta_3={\rm Li}_3(1)=1.2020569...$\ .\\
For large $y = s/(2\mu)^2$ one obtains
\begin{equation}
\int_{1/y}^1 \frac{{\rm d}\lambda}{\lambda} F(\lambda) = 
\left[ \frac{1}{3} \ln^3y - \frac{3}{2} \ln^2y + 
\left(-2\zeta_2+5\right)\ln y+\left(6\zeta_3+3\zeta_2-19/2\right) \right]\,.
\label{equ.5}
\end{equation}
The second integral is dominated by low $s'$. The function $R(s')$
approaches rapidly the asymptotic value $R(\infty)$ and the remaining
integral can be extended to infinity. Only small values of $s'/s$ contribute
to the integral, and $F(\lambda)$ can be approximated by its behaviour for
small $\lambda$
\begin{equation}
F(\lambda) \stackrel{\lambda \to 0}{\longrightarrow} \ln^2\lambda +
3 \ln\lambda - 2\zeta_2 + 5 \,.
\label{equ.6}
\end{equation}
Hence
\begin{equation}
\int_{(2\mu)^2}^s \frac{{\rm d}s'}{s'} \left[R(s') - R(\infty)\right]
F(s'/s) \longrightarrow \int_0^1 \frac{{\rm d}x}{x} \left[
R\left((2\mu)^2/x\right) - R(\infty)\right] \left[
\ln^2xy - 3\ln xy - 2\zeta_2 +5 \right].
\label{equ.7}
\end{equation}
Combining the two contributions one arrives at
\begin{eqnarray}
\frac{\Gamma_{l \bar l had}}{\Gamma_{l \bar l}} &=&
\frac{1}{6} \left(\frac{\alpha}{\pi}\right)^2 \biggl\{
R(\infty) \left[ \frac{1}{3} \ln^3y - \frac{3}{2} \ln^2y + 
\left(-2\zeta_2+5\right)\ln y+\left(6\zeta_3+3\zeta_2-19/2\right)
\right]+\nonumber\\
& & R_0 \left(\ln^2y-3\ln y-2\zeta_2+5\right) +
R_1 \left(2\ln y-3\right) + 2 R_2 \biggr\}\,,
\label{equ.8}
\end{eqnarray}
where the moments $R_n$ are defined through
\begin{equation}
R_n = \int_0^1 \frac{{\rm d}x}{x} \frac{\ln^nx}{n!}
\left(R(4m^2/x)-R(\infty)\right)\,.
\label{equ.9}
\end{equation}
This result can be used to predict the inclusive hadronic rate, if we
employ $R(s)$ as provided from experiment \cite{Burkhardt}.
\begin{figure}
\label{fig2}
\begin{center}
\leavevmode
\epsffile[0 40 377 243]{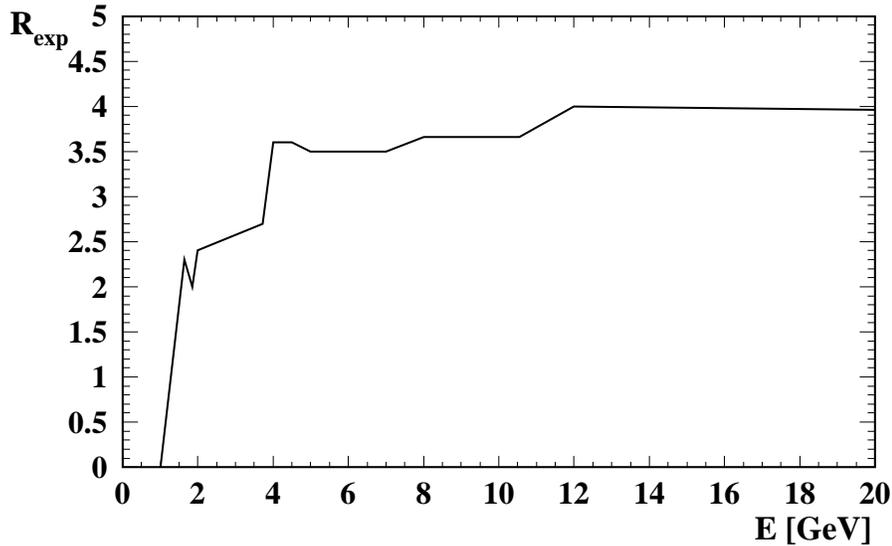}
\caption{{\em Parametrisation of $R(s)$ in the threshold region, without
resonances, as provided by {\protect\cite{Burkhardt}}.}}
\end{center}
\end{figure}
A convenient approach adopted to the experimental procedures is to decompose
the final states into continuum (Fig.2) and resonance
contributions. For the continuum one finds
\begin{equation}
R(\infty)=11/3\,,\  R_0=-13.240\,,\  R_1=20.875\,,\  R_2=-3.587\,.
\label{equ.10}
\end{equation}
For the two pion ($\widehat{=} \rho$) contribution the
parametrisation derived in \cite{KS} (set 1) is employed
which leads to the following moments:
\begin{equation}
R(\infty)=0\,,\  R_0=4.319\,,\  R_1=-8.422\,,\  R_2=8.547\,.
\label{equ.11}
\end{equation}
For the remaining resonances the narrow width approximation can be applied
\begin{equation}
R(s) = \frac{9\pi \Gamma_{e^+e^-} M_R}{\alpha^2} \delta(s-M_R^2)\,.
\label{equ.12}
\end{equation}
The input parameters are listed in table \ref{table1}, together
with the predicted relative rate.
\begin{table}
\caption{{\em Input parameters used for the narrow resonances and
relative rates of hadronic radiation.}}
\begin{center}
\begin{tabular}{|c||c|r|r|} \hline
Channel & $\Gamma_{e^+e^-}\ [{\rm keV}]$ & $M\ [{\rm MeV}]$ & Relative Rate \\
\hline \hline
$2\pi$ & & & $2.55\times 10^{-4}$ \\ \hline
Continuum & & & $2.86\times 10^{-4}$ \\ \hline
$\omega$ & $0.60$ & $781.95$ & $2.34\times 10^{-5}$ \\ \hline
$\phi$ & $1.37$ & $1019.41$ & $3.56\times 10^{-5}$ \\ \hline
$\psi$ & $5.36$ & $3096.93$ & $2.25\times 10^{-5}$ \\ \hline
$\psi'$ & $2.14$ & $3686.00$& $6.55\times 10^{-6}$ \\ \hline
$\Upsilon$ & $1.34$ & $9460.32$ & $5.85\times 10^{-7}$ \\ \hline
\end{tabular}
\end{center}
\label{table1}
\end{table}
In total we find the following result
\begin{equation}
\frac{\Gamma_{l \bar l had}}{\Gamma_{l \bar l}} = 6.30 \times 10^{-4}\,.
\label{equ.13}
\end{equation}
\begin{table}
\caption{{\em Contributions of different energy ranges to the relative rate
\protect{$\Gamma_{l \bar l had}/\Gamma_{l \bar l}$}.}}
\begin{center}
\begin{tabular}{|c||c|} \hline
Energy range & Contribution
to $\Gamma_{l \bar l had}/\Gamma_{l \bar l}$\\ \hline \hline
$0\le m_{had}\le 2$ GeV & $3.88\times 10^{-4}$ \\ \hline
$2\ {\rm GeV}\le m_{had}\le 3$ GeV & $5.87\times 10^{-5}$ \\ \hline
$3\ {\rm GeV}\le m_{had}\le 10$ GeV & $1.52\times 10^{-4}$ \\ \hline
$10\ {\rm GeV}\le m_{had}\le M_Z$ & $3.16\times 10^{-5}$ \\ \hline
\end{tabular}
\end{center}
\label{table2}
\end{table}
It should be reiterated that this contribution corresponds to
configurations where the invariant mass of the lepton pair is typically
large and the mass of the hadronic system small. This is evident from
table \ref{table2} where the contribution for $m_{had}\ge 10 \ {\rm GeV}$
is also shown separately.\\
The numbers listed in tables \ref{table1} and
\ref{table2} are difficult
to compare with those from \cite{vdB}. There the calculation was
performed numerically for a $q\bar q l\bar l$ final state with a fixed
quark mass of $300$ MeV. For $l=e$ and to lesser extend for $l=\mu$
their result is necessarily dominated by final states with large
hadronic mass, that is a multi--hadron final state plus a lepton pair
of low invariant mass. It seems experimentally difficult to isolate
these events among the multi--hadron states, which occur with a
relative rate below $10^{-3}$. It is however an interesting exercise
to apply (\ref{equ.8}) and the formulas discussed below
to $l\bar l q\bar q$ production where again the hadrons are radiated
off the lepton. These can experimentally isolated by considering
events with large invariant mass of the lepton pair.
Adopting the quark mass values of \cite{vdB}, viz.
$m_u=m_d=m_s=300 \ {\rm MeV}$, $m_c=1.5 \ {\rm GeV}$
and $m_b=4.5 \ {\rm GeV}$ one predicts
\begin{eqnarray}
\frac{\Gamma_{l\bar l u\bar u}}{\Gamma_{l\bar l}}&=&
4\frac{\Gamma_{l\bar l d\bar d}}{\Gamma_{l\bar l}}=
4\frac{\Gamma_{l\bar l s\bar s}}{\Gamma_{l\bar l}}=2.256\times
10^{-4}\,,\nonumber\\
\frac{\Gamma_{l\bar l c\bar c}}{\Gamma_{l\bar l}}&=&5.259\times
10^{-5}\,,\nonumber\\
\frac{\Gamma_{l\bar l b\bar b}}{\Gamma_{l\bar l}}&=&2.858\times
10^{-6}\,,\nonumber\\
\frac{\Gamma_{l\bar l had}}{\Gamma_{l\bar l}}&=&3.938\times 10^{-4}\,.
\label{equ.14a}
\end{eqnarray}
We find that this model which has been used in \cite{vdB}, underestimates
the rate for radiation by a factor $0.63$. An experimental analysis
based on this model could, therefore, lead to drastically wrong conclusions.
The reason for the discrepancy is fairly evident: The events are dominated
by hadronic states with low invariant mass, where the parton model
provides a poor approximation. This is reflected in a strong sensitivity
of the results \pagebreak[4] towards the choice for $m_q$.\\
\noindent
{\em Four lepton final states.}
The same formalism can also be applied to purely leptonic final states,
using the analytical results for the moments\footnote{
The result for the total cross section in
leading logarithmic approximation can be found in \cite{Khoze}.}
\cite{KKKS}
\begin{eqnarray}
R(\infty) &=& 1\,, \quad R_0 = \ln 4 - \frac{5}{3}\,, \nonumber\\
R_1 &=&  \frac{1}{2} \ln^2 4 - \frac{5}{3} \ln 4 + \left[ \frac{28}{9}
- \zeta_2 \right]\,, \nonumber\\
R_2 &=& \frac{1}{6} \ln^3 4 - \frac{5}{6} \ln^2 4 + \left[
\frac{28}{9} - \zeta_2 \right] \ln 4 + 2 \zeta_3 + \frac{5}{3} \zeta_2
- \frac{164}{27}\,.
\label{equ.14}
\end{eqnarray}
The numerical results for $\Gamma_{e^+e^- l \bar l} / 
\Gamma_{e^+e^-}$ are $3.319 \times 10^{-4}$ for $l=\mu$
and $3.214 \times 10^{-5}$ for $l=\tau$.
It should be reiterated that this result refers to final states where
the $l\bar l$ is radiated off the $e^+e^-$ pair and hence is
dominated by $l\bar l$ with low invariant mass.
Eq.\ref{equ.8} can be also used to predict the rate for lepton pair
radiation if a cut $s_{l^+l^-} > s_0 \gg 4 m_l^2$ is
employed. In this case one may
simply put $R_0 = R_1 = R_2 = 0$ and $y = s/s_0$\,. Characteristical
examples are $\Gamma_{e^+e^- l^+l^-}/\Gamma_{e^+e^-}=5.321\times 10^{-4} /
1.268\times 10^{-4} / 8.716\times 10^{-6}$ for
$\sqrt{s_0}=0.1 / 1 / 10 \ {\rm GeV}$ respectively.\\

\noindent
{\em Secondary radiation of heavy quarks.}
Replacing the factor $(\alpha / \pi)^2$ by $(\alpha_s / \pi)^2\,2/3$,
the perturbative result of eq.\ref{equ.8} combined with eq.\ref{equ.14}
can be applied to predict the rate for secondary production of heavy
quarks in $Z \to q\bar q$ events, where $q$ denotes a light quark. For
$m_b=4.5/4.8/5.0$ GeV one predicts a relative rate of $1.545\times 10^{-3}/
1.391\times 10^{-3}/1.301\times 10^{-3}$. The corresponding numbers for
secondary charm production $(m_c=1.3/1.5/1.8\ {\rm GeV})$ are significantly
larger and amount to $8.339\times 10^{-3}/7.107\times 10^{-3}/5.729\times
10^{-3}$ respectively. These rates should be well observable with present
day statistics.\\

\noindent
{\em Combination with virtual corrections.}
The third and second power of the logarithm of $y$ cancel after combining
real radiation of hadrons (or leptons) with the corresponding virtual
corrections to the form factor \cite{KKKS}
\begin{eqnarray}
{\rm Re}F^{(4)} &=& R(\infty) \left\{ -\frac{1}{36}\ln^3y+\frac{1}{8}
\ln^2y+\left[\frac{1}{6}\zeta_2-\frac{7}{24}\right]\ln y-\frac{1}{4}
\zeta_2+\frac{5}{16}\right\} + \nonumber\\
& & R_0\left[-\frac{1}{12}\ln^2y+\frac{1}{4}\ln y+\frac{1}{6}\zeta_2-
\frac{7}{24}\right]+R_1 \left(-\frac{1}{6}\ln y+\frac{1}{4}\right) - 
\frac{1}{6}R_2\,,
\label{equ.17}
\end{eqnarray}
and one obtains for the correction to the total rate
\begin{eqnarray}
\frac{\delta \Gamma^{(4)}}{\Gamma} &=& 2 \left(\frac{\alpha}{\pi}\right)^2
{\rm Re}F^{(4)}+
\frac{\Gamma_{e^+e^- l \bar l}}{\Gamma_{e^+e^-}}\nonumber\\
\hbox{} &=&\left(\frac{\alpha}{\pi}\right)^2\left\{R(\infty)
\left[\frac{1}{4}\ln y-\frac{23}{24}+\zeta_3\right]+\frac{1}{4}R_0\right\}\,.
\label{equ.18}
\end{eqnarray}
The coefficient of the remaining logarithm can be easily understood:
The lowest order correction (from real and virtual photon radiation)
is given by a factor $(1+\frac{3}{4}\frac{\alpha}{\pi})$. The fine
structure constant $\alpha$ can be replaced by the running
$\alpha(s)=\alpha/(1-\Pi(s))$.
The vacuum polarisation from hadronic (or leptonic) intermediate states
can be expressed in terms of $R$ through
\begin{eqnarray}
{\rm Re} \Pi(s) &=& \frac{\alpha}{\pi} \frac{s}{3} \int_{4\mu^2}^{\infty}
\frac{{\rm d}s'}{s'} \frac{R(s')}{s-s'}\nonumber\\
\hbox{}&\approx& \frac{\alpha}{\pi}\left(\frac{1}{3}R(\infty)\ln
\frac{s}{4\mu^2}+\frac{1}{3}R_0\right)\,,
\label{equ.19}
\end{eqnarray}
and one obtains for the complete correction
(without ${\cal O}(\alpha^2)$ photonic terms)
\begin{equation}
1+\frac{3}{4}\frac{\alpha(s)}{\pi}+\frac{1}{6}\left(\frac{\alpha}{\pi}
\right)^2 \left(6\zeta_3-\frac{23}{4}\right)\,.
\label{equ.20}
\end{equation}
An amusing test of (\ref{equ.18}), combined with the moments
from (\ref{equ.14}), is provided by the results of \cite{Chet}.
If one replaces the $SU(3)$ factors in this paper by the corresponding
abelian coefficients, and, furthermore, relates
the $\overline{MS}$ coupling $\alpha_{\overline{MS}}$ to the
conventionally defined fine structure constant
\begin{equation}
{\overline \alpha} = \alpha \left( 1 + \alpha \frac{1}{3} \ln
\frac{\mu^2}{m^2} \right)
\label{equ.21}
\end{equation}
one obtains agreement between the two results.\footnote{We acknowledge
a helpful discussion with K. Chetyrkin on this comparison.}\\
\vskip 1.5cm
\sloppy
\raggedright
\def\app#1#2#3{{\it Act. Phys. Pol. }{\bf B #1} (#2) #3}
\def\apa#1#2#3{{\it Act. Phys. Austr.}{\bf #1} (#2) #3}
\def\lhc{Proc. LHC Workshop, CERN 90-10}
\def\npb#1#2#3{{\it Nucl. Phys. }{\bf B #1} (#2) #3}
\def\plb#1#2#3{{\it Phys. Lett. }{\bf B #1} (#2) #3}
\def\prd#1#2#3{{\it Phys. Rev. }{\bf D #1} (#2) #3}
\def\pR#1#2#3{{\it Phys. Rev. }{\bf #1} (#2) #3}
\def\prl#1#2#3{{\it Phys. Rev. Lett. }{\bf #1} (#2) #3}
\def\prc#1#2#3{{\it Phys. Reports }{\bf #1} (#2) #3}
\def\cpc#1#2#3{{\it Comp. Phys. Commun. }{\bf #1} (#2) #3}
\def\nim#1#2#3{{\it Nucl. Inst. Meth. }{\bf #1} (#2) #3}
\def\pr#1#2#3{{\it Phys. Reports }{\bf #1} (#2) #3}
\def\sovnp#1#2#3{{\it Sov. J. Nucl. Phys. }{\bf #1} (#2) #3}
\def\jl#1#2#3{{\it JETP Lett. }{\bf #1} (#2) #3}
\def\jet#1#2#3{{\it JETP Lett. }{\bf #1} (#2) #3}
\def\zpc#1#2#3{{\it Z. Phys. }{\bf C #1} (#2) #3}
\def\ptp#1#2#3{{\it Prog.~Theor.~Phys.~}{\bf #1} (#2) #3}
\def\nca#1#2#3{{\it Nouvo~Cim.~}{\bf #1A} (#2) #3}
{\elevenbf\noindent  References \hfil}
\vglue 0.4cm

\end{document}